\documentstyle[psbox]{EAYAM2006}
\begin{document}

\title{
Open clusters in a dispersing molecular cloud \\
{\large\sf } 
}

\author{
Hui-Chen Chen$^1$, Chung-Ming Ko$^1$$^,$$^2$
\\[12pt]  
%
$^1$  Institute of Astronomy, National Central University, Taiwan \\
$^2$  Department of Physics and Center for Complex System, National Central University, Taiwan \\
%
{\it E-mail(HC):huichen@astro.ncu.edu.tw} }

\abst{
Star clusters are formed in molecular clouds which are believed 
to be the birth places of most stars. From recent observational data, 
Lada \& Lada(2003) estimated that only 4 to 7\% of the 
proto-clusters have survived. Many factors could cause this high infant 
mortality. Galactic tidal forces, close encounters with molecular clouds and 
shock heating are among the possible causes but they have a longer timescale than 
typical lifetime of molecular clouds. Another possible 
reason is mass loss in very beginning of cluster evolution in the form of UV radiation, stellar winds 
or supernova explosions. Mass loss is the main factor we 
study in this work by using N-body simulations. We find that 
most proto-clusters survive for more than 40 Myr even when the mass loss rate
 is high. 
}


\maketitle \thispagestyle{empty}

\section{Introduction}
Stellar clusters are among the most interesting objects in astronomy. 
We are interested in how they form, how they evolve and how they die.
Theories and observations indicate that most stars are not born independently 
but in stellar clusters or stellar associations
which are formed in molecular clouds. Due to the limits of
observational techniques, we do not know in detail
the relation between the clusters and the parent molecular clouds.
It is believed that stars are born in clusters and become field
stars after the clusters disassociate.

Recently, near
infrared observational data (2MASS, Two Micron All Sky Survey
project at IPAC/Caltech) have shown that the number of embedded clusters is much
higher than the number of optical clusters for which the parent clouds
have already dissipated and that the survival probability for proto-clusters is about 4 to 7\% 
(Lada and Lada 2003). This implies that clusters are likely to
be disrupted before the clouds are dissipated completely. 
From the time the clusters are born, the surrounding environments keep dissolving 
them. Galactic tidal forces, close encounters 
with giant molecular clouds, shock heating and mass loss 
by massive member stars (Boily and Kroupa 2003) are possible 
dissolving mechanisms that operate during the clusters' lifetime. Nonetheless, most of them have a longer timescale 
than the lifetime of molecular clouds, which is less than 10 Myr (Williams et al 1999).
Naively speaking, the mechanism which works within the lifetime of the clouds should 
be the main reason of this low survival probability. In this work, 
we focus on the effects of mass loss of the cloud in the early evolution of the system.

In the beginning, the clusters are 
bound to the molecular clouds. 
As the clouds dissipate, the binding energy from the cloud decreases 
and the stellar systems become out of equilibrium. Once out of equilibrium, they may expand or 
dissociate completely.

\section{Simulations}

Due to the very different size and density between the clouds and the clusters, 
it is not easy to simulate stars and gas clouds together.
In this work, we
adopted the NBODY2 code developed by Aarseth (Aarseth
2001) to study the dynamical evolution of the clusters.

\subsection{Model for cloud dispersion}
The initial distribution of the clouds is represented by a 
potential energy in the form of the Plummer model(1911),

\begin{equation}
\Phi_{P}=\frac{-GM_b}{\sqrt{r^2+a^2}},
\end{equation}
where {\it M$_b$} is the mass of the cloud, {\it a} is the length
scale of the potential and G is the gravitational constant. To
model the dispersion of the cloud we allow the potential energy to evolve 
in time according to 
\begin{equation}
a=a_0^{\alpha t},
\end{equation}
where {\it a$_0$} is the initial length scale of the cloud and
$\alpha$ is the dispersion rate of the cloud. The mass of the 
cloud, {\it M$_b$} remains constant as the length scale 
increases with time {\it t}. The potential well of this system 
tends to become shallower in time so that the velocity of some stars can 
exceed the escape velocity and the stars can run away.

\subsection{Initial conditions}
The stars are initially distributed according to a Plummer distribution both in physical 
positions and velocities which are required to achieve virial equilibrium. 
Note that the fact the clusters are in virial equilibrium does not mean that they are also 
in dynamical equilibrium(Goodwin 1977). The steps for
generating the initial conditions are as followed:
\begin{itemize}
  \item generate a cluster with a Plummer distribution and
with a size of about 1 pc.
  \item put the cluster into a molecular cloud, represented as a Plummer
potential energy, in the same center of mass and set the dispersion rate $\alpha$ 
of the cloud to zero, which means that the molecular cloud 
will not change with time.
  \item run the code until the cluster is in a quasi-steady state.
  \item use this quasi-steady state as initial condition for the simulations, 
  set the dispersion rate $\alpha$ to be greater than zero 
  and let the evolution start.
\end{itemize}

In our simulations we study three free parameters.
First, the mass of the cloud, {\it M$_b$}, is a multiple of the stellar
cluster mass {\it M$_{c}$}, from 1 to 10, resulting in a star formation efficiency, 
\begin{equation}
\eta=\frac{M_{c}}{M_{c}+M_b},
\end{equation}
of 50 to 9\%, respectively.
Second, the compactness of the cloud, which is controlled by the initial length
scale of the potential energy, {\it a$_0$}, runs from 0.25 pc to 2.5 pc.
Third, the dispersion rates of the cloud, $\alpha$, are 0.05,
0.1 and 0.2 Myr$^{-1}$. We performed three hundred runs which are described in 
Table 1 for the single mass cases. Table 2 lists the parameters of three typical cases 
which we select for further analysis. 
\\

In each simulation, we use 2500 particles and consider two
kinds of mass function: a single mass distribution and 
a mass function with slope is -3 (where Salpeter is -2.35)
 from 0.15 to 12 M$_\odot$.
\\

\begin{table}[htdp]
\caption{parameters}
\begin{center}
\begin{tabular}{l c c c c c c c}
Run & 1 & 2 & 3 & $\cdots$ & 8 & 9 & 10 \\ \hline
 {\it M$_b$} [{\it M$_{c}$}] &  1 & 2 & 3 & $\cdots$ & 8 & 9 & 10 \\
 $\eta$ [\%] & 50 & 33 & 25 & $\cdots$ & 11 & 10 & 9 \\
 \\
 Run & A & B & C & $\cdots$ & H & I & J \\ \hline
 {\it a$_0$} [pc] & 0.25 & 0.5 & 0.75 & $\cdots$ & 2 & 2.25 & 2.5\\
 \\
 Run & a & b & c \\ \hline
 $\alpha$  [Myr$^{-1}$] & 0.2 & 0.1 & 0.05 \\
\end{tabular}
\end{center}
\label{parameters}
\end{table}%

\begin{table}[htdp]
\caption{case study}
\begin{center}
\begin{tabular}{c c c c}
Run & {\it M$_b$} & {\it a$_0$} & $\alpha$ \\
 & [{\it M$_{cluster}$}] & [pc] & [Myr$^{-1}$] \\ \hline
1Aa & 1 & 2.5 & 0.2 \\
5Ea & 5 & 1.25 & 0.2 \\
10Ja & 10 & 0.25 & 0.2 \\ \hline
\end{tabular}
\end{center}
\label{modes}
\end{table}%

\section{Results}

\subsection{Case study}
In order to investigate parameter space ({\it M$_b$}, {\it a$_0$}), 
we perform a few hundred runs for this work. 
To illustrate the main results, we present three typical cases which are listed in Table 2: 
(1Aa) {\it M$_b$}=1 stellar cluster
mass, {\it a$_0$}=2.5 pc, less massive and loose cloud. (5Ea) {\it
M$_b$}=5 stellar cluster mass, {\it a$_0$}=1.25 pc, intermediate property 
cloud. (10Ja) {\it M$_b$}=10 stellar cluster mass, {\it a$_0$}=2.5
pc, massive and compact cloud. Fig.1 shows the number of stars varying in time while 
Fig.2 shows the Lagrangian radii (the radii which contain different percentages of the cluster mass) for 
the three cases.
We find that: \\

{\it Case 1Aa}:
the masses of the cloud and of the cluster are comparable, which means that the binding energy 
of this system is not high in the beginning. The structure of the cluster was rarely affected by 
the dispersion of the cloud which can be seen in Fig.1(a) and 
fig.2(a). In Fig.1(a), the number 
of stars decreases slowly in time, and it does not change significantly even after 250 Myr. 
In the inner region, 0.5 pc, the number of stars increases slightly due to the cluster contraction. 
Fig.2(a) indicates that only the 90\% Lagrangian radius 
increases by a factor of two while the 50 and 70\% Lagrangian radii increase slightly. 
Similarly, the 10 and 30\% Lagrangain radii show a small increase. 
In this case, the cluster expands in the 
outer part but contracts in the inner part.
More than 2000 stars remain in the inner 2 pc even after 
250 Myr. We therefore argue that the cluster is still bound.
\\

{\it Case 5Ea}: 
the star formation efficiency is 13\% and the binding energy is much higher than Case 1Aa. 
The number of stars decrease quickly in the fisrt 5 Myr, 
but becomes stable after 10 Myr, as shown in Fig.1(b).
Fig.2(b) indicates that all the Lagrangian radii increase in the beginning but stop at about 10 Myr, 
with the except 90\% Lagrangian radius, which expands for a longer time 
but also becomes stable after 35 Myr. 
Therefore, the cluster expends due to the dispersion of the cloud and becomes looser: the number 
density reduces to half within a 2 pc radius.
\\

{\it Case 10Ja}:
in this very low star formation efficiency case, shown in Fig.1(c) and Fig.2(c), 
the number of stars drops to zero in 10 Myr and the Lagrangian radii increase.
The expansion continues till the end of the simulation.
The cluster is disrupted within a few Myr. 
\\
\subsection{Systems with equal mass}
Fig.3(a) shows the final number of stars within 2 pc in the 
runs with a single mass distribution. In larger and less massive clouds, 
the final number of stars
within 2 pc is high. The number of stars decreases as {\it a$_0$}
becomes smaller and {\it M$_b$} becomes larger. 
In fig.3(b), the final 50\% Lagrangian radius 
is high for smaller {\it a$_0$} and larger {\it M$_b$}. 
The final radius decreases as {\it a$_0$} becomes larger and
{\it M$_b$} becomes smaller. The density for 500 stars in 2 pc
radius is about 15/pc$^3$, which is much higher than the field star
density, 0.1/pc$^3$. The 500 contour line in the final number plot
and the 5 pc contour line in the final radius plot match perfectly.
We conclude that the clusters survived at least beyond this line.

\subsection{Systems with mass funcion}
Observations show that there is an initial mass function in stellar
clusters. Here we generate the clusters by taking -3.0 as the mass
function slope(where the Salpeter's slope is -2.35), and the
mass range is from 0.15 to 12 M$_\odot$.
In every simulation, we also consider 2500 stars as in the cases of the single mass model. 
Fig.4 shows that 
the trend of the final number of stars within 2 pc and the final
radius of the inner 50\% of stars are similar to the simulations for
the single mass distribution. Also, the 500 contour line in the final
number plot and the 5 pc contour line in the final radius plot
match each other.

\subsection{Dispersion rates}
There are three dispersion rates of the cloud we adopted for the single mass model, 
0.05, 0.1 and 0.2 Myr$^{-1}$.
It seems that even in the $\alpha$=0.05 Myr$^{-1}$ cases, for which the e-fold time is 4 times smaller 
than 0.2 Myr$^{-1}$, the result is quite similar. 
The larger dispersion rates will speed the evolution, but not much. 
This supports the idea that the survival probability is high during the cloud dissipation 
even with a very high dissipation rate.

\begin{figure*}[fig1]
\centering
\psbox[xsize=12 cm,rotate=l] {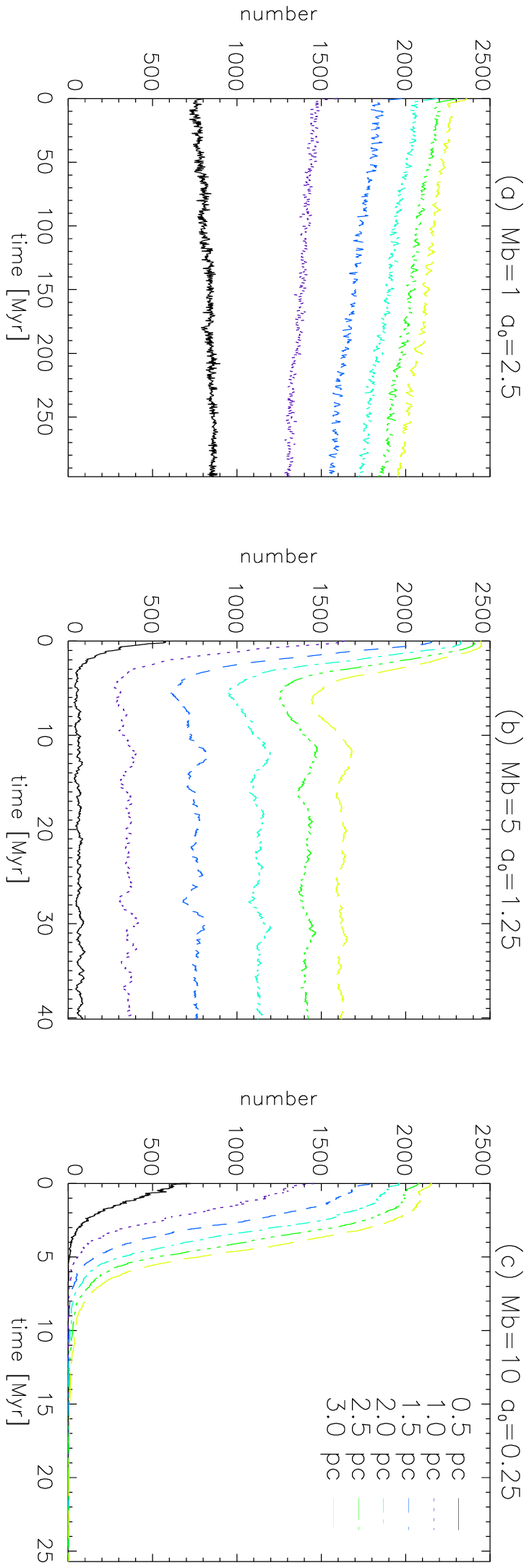} \caption{
Number of stars in fix radii at $\alpha$=0.2 Myr$^{-1}$. 
(a) Case 1Aa,  (b) Case 5Ea and (c) Case 10Ja.}
\psbox[xsize=12 cm,rotate=l] {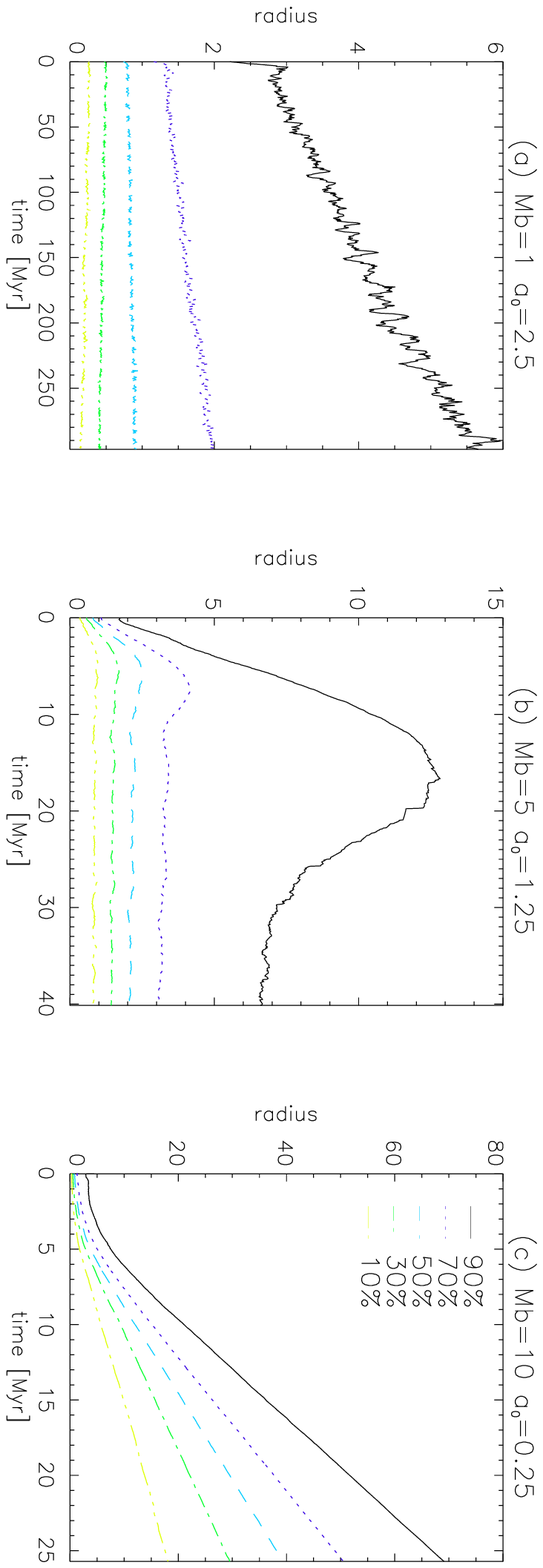} \caption{
Lagrangian radii at $\alpha$=0.2 Myr$^{-1}$. 
(a) Case 1Aa,  (b) Case 5Ea and (c) Case 10Ja.}

\centering
 \psbox[xsize=12 cm,rotate=l] {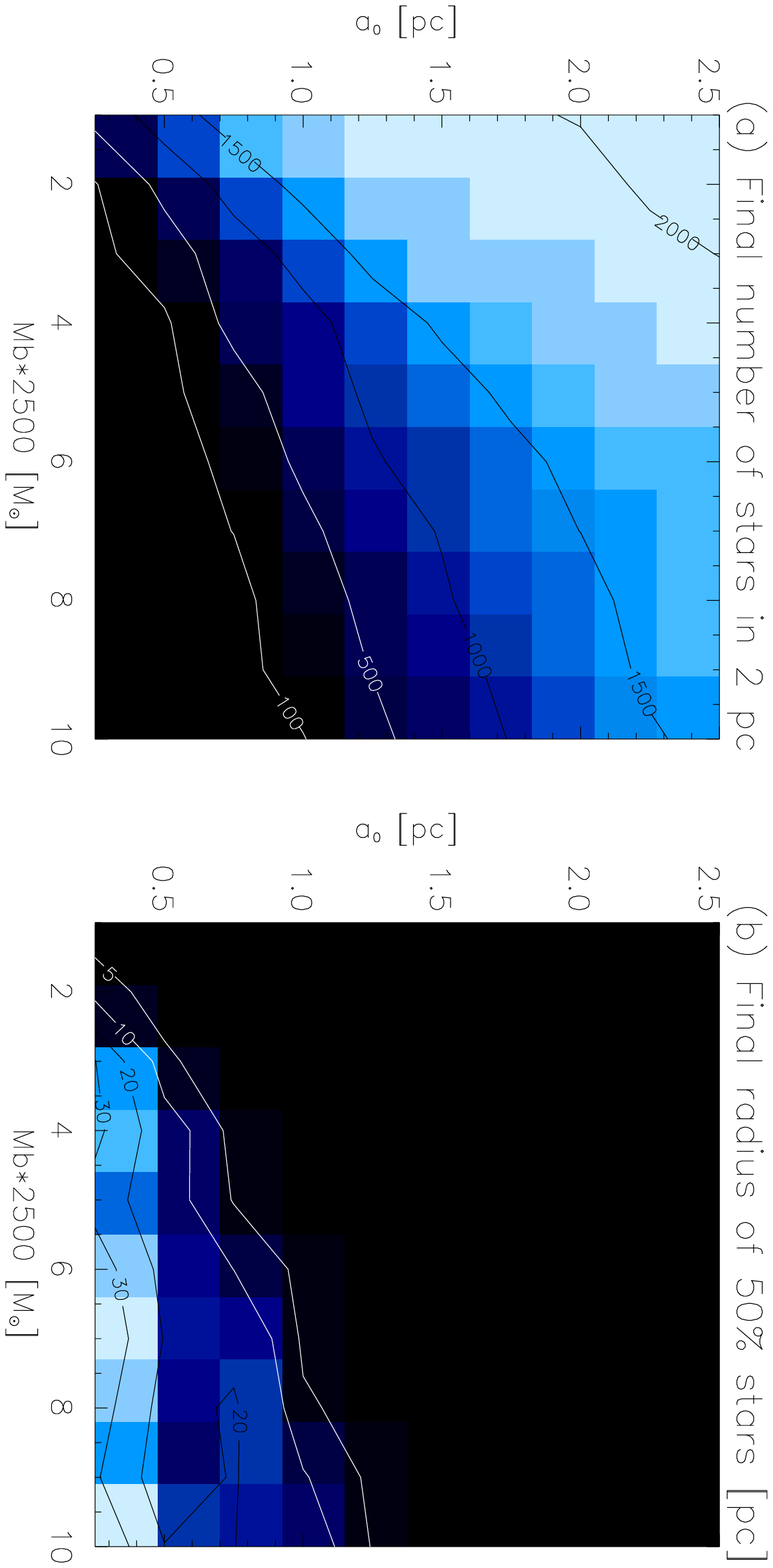}
\caption{
Result for single mass distribution at $\alpha$=0.2
Myr$^{-1}$. (a) final number of stars within 2 pc, (b) final
50 \% Lagrangian radius [pc]. }

\psbox[xsize=12 cm,rotate=l] {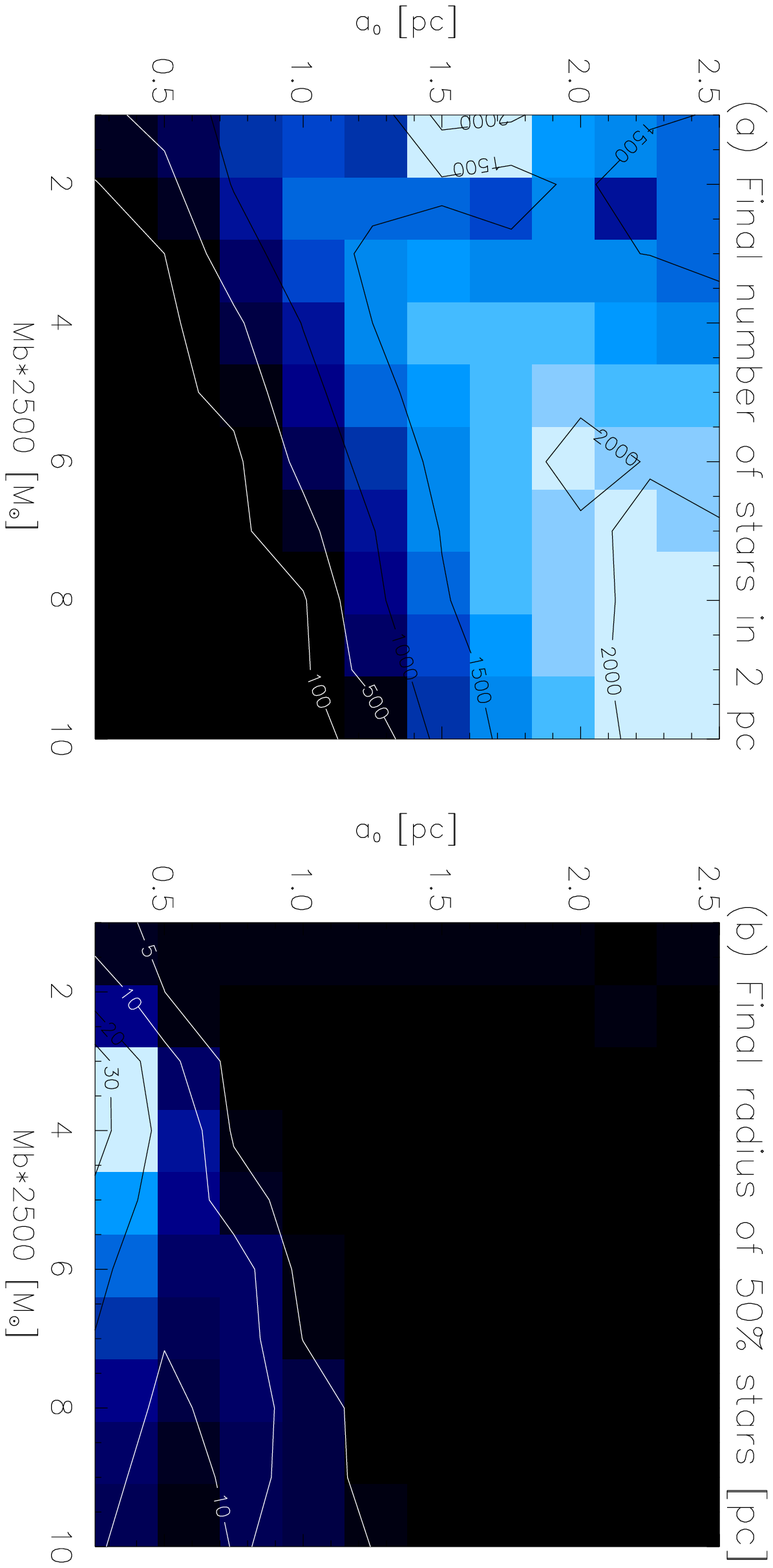} \caption{
Result for stars with a IMF slope=3.0 at $\alpha$=0.2 Myr$^{-1}$. 
(a) final number of stars within 2 pc, (b) final
50 \% Lagrangian radius [pc].}
\end{figure*}

\section{Conclusions}
The proto-clusters can be disrupted by mass loss in the 
early stages of evolution. In this work, we studied this behavior by means of N-body 
simulations. The mass loss tends to expand or even disassociate 
the clusters due to the decrease in binding energy. 
The survival probabilities are high both when the star formation 
efficiency is high and when the molecular cloud is loose. 
More than 75\% of the clusters retain a core, with a 
number density higher than 15/pc$^3$ after few hundred Myr. 
Different dispersing rates for the cloud provide similar results and even 
the largest rate does not disrupt all the clusters. Systems with and without 
the initial mass function have different final densities but agree 
with each other very well. We conclude that the infant mortality 
should be low if the proto-clusters are bound from the beginning.

\vspace{1pc} \noindent
\section*{Acknowledgements}
The authors appreciate Alessia Gualandris for her kindly and very helpful discussions.
This work was supported in part by the National
Science Council, Taiwan under the Grants
NSC-093-2112-M-008-017 and NSC-94-2112-M-008-018.

\section*{References}
\re Aarseth, S.J. 2001 New Astronomy, 6, 277 \re Boily, C.M. and
Kroupa, P.  2003 MNRAS, 338, 665 \re Boily, C.M. and Kroupa, P.
2003 MNRAS, 338, 673 \re Goodwin, S.P. 1997 MNRAS, 284, 785 \re
Lada, C..J. and Lada, E.A.  2003 ARA\&A, 41, 57
\end{document}